\documentclass{article}
\usepackage[T1]{fontenc} 
\usepackage[utf8]{inputenc} 
\usepackage{ismir,amsmath,cite,url}
\usepackage{graphicx}
\usepackage{color}
\usepackage{multirow}
\usepackage{multicol}
\usepackage{rotating}
\usepackage{makecell}
\usepackage{tabularx}
\usepackage{adjustbox}
\usepackage{booktabs}
\usepackage{graphicx}
\usepackage{subcaption}

\usepackage{amsfonts}
\usepackage{enumitem}

\title{Towards Zero-Shot Amplifier Modeling: One-to-Many Amplifier Modeling via Tone Embedding Control}

\multauthor
{Yu-Hua Chen$^{1,3}$ \hspace{1cm} Yen-Tung Yeh$^{2,3}$ \hspace{1cm} Yuan-Chiao Cheng$^3$} { \bfseries{Jui-Te Wu$^3$ \hspace{1cm} Yu-Hsiang Ho$^3$ \hspace{1cm} Jyh-Shing Roger Jang$^1$ \hspace{1cm} Yi-Hsuan Yang$^2$}\\
$^1$ Graduate Institute of Networking and Multimedia, National Taiwan University, Taiwan\\
$^2$  Department of Electrical Engineering, National Taiwan University, Taiwan\\
$^3$ Positive Grid, Taiwan
}

\def\authorname{Yu-Hua Chen, Yen-Tung Yeh, Yuan-Chiao Cheng, Jui-Te Wu, Yu-Hsiang Ho, Jyh-Shing Roger Jang, and Yi-Hsuan Yang}

\begin{document}

\maketitle
\begin{abstract}
Replicating analog device circuits through neural audio effect modeling has garnered increasing interest in recent years. Existing work has predominantly focused on a one-to-one emulation strategy, modeling specific devices individually. In this paper, we tackle the less-explored scenario of one-to-many emulation, utilizing conditioning mechanisms to emulate multiple guitar amplifiers through a single neural model.
For condition representation, we use contrastive learning to build a tone embedding encoder that extracts style-related features of various amplifiers, leveraging a dataset of comprehensive amplifier settings. Targeting zero-shot application scenarios, we also examine various strategies for tone embedding representation, evaluating referenced tone embedding against two retrieval-based embedding methods for amplifiers unseen in the training time. Our findings showcase the efficacy and potential of the proposed methods in achieving versatile one-to-many amplifier modeling,  contributing a foundational step towards zero-shot audio modeling applications.

\end{abstract}

\section{Introduction}
\label{sec:intro}


Neural audio effect modeling, the task of simulating analog circuitry and digital audio effects using neural networks, has garnered significant interest driven by advances in deep learning~\cite{damskagg2019deep,wright2019real, wright2020real, martinez2020deep, juvela2023end, comunita2023modelling, yin2024modeling,chen24dafx,yeh24dafx}. 
Various network architectures have been proposed for emulating different effect pedals and guitar amplifiers (amps).
Through modeling nonlinearities, harmonic distortions, and transient responses inherent to analog circuitry, neural models offer an alternative to their physical counterparts. 
Such models enable widespread applications in 
automatic mixing~\cite{steinmetz2021automatic, martinez2022automatic, koo2023music}, audio style transfer~\cite{mimilakis2020one, steinmetz2022style} and beyond, contributing to new 
music production and sound design workflows. 

\begin{figure}[t]
 \centerline{{
 \includegraphics[width=0.9\columnwidth]{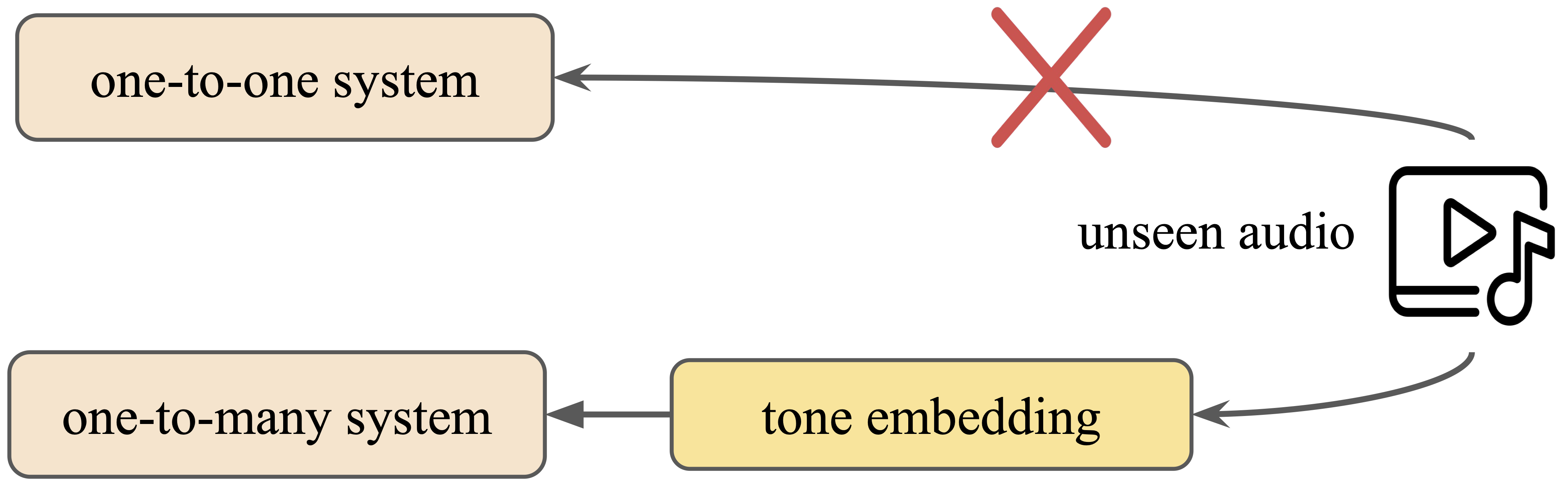}}}
 \caption{A one-to-one approach cannot emulate an unseen audio effect. In contrast, the proposed one-to-many approach can achieve zero-shot modeling by using a tone embedding encoder that turns a reference audio example of that effect into a conditioning input at inference time.}
 \label{fig:intro}
\end{figure}


Real-world audio effect pedals or amplifiers are known for their rich acoustic diversity, leading to different ``tones.'' There are multiple types of pedals (e.g., compression, EQ, distortion, reverb, modulation), possessing different characteristics. Even pedals of the same effect type can sound fairly differently due to different implementations. A user can further adjust the device parameters via tuning the associated knobs (e.g., a gain knob) to shape the tone. Moreover, it is common to interconnect effect pedals in various orders and forms to constitute an ``effect chain,'' collectively creating a unique tone. 
A guitar amplifier can also be viewed as an effect chain as there are pre-amp, tone stack, power amp, and cabinet components arranged in specific order that vary across brands and underlying  circuitries. 
The rich diversity of tones represents a central challenge for neural audio effect modeling.

While exciting progress is being made, 
to simplify the task,
most prior research has  concentrated on the \emph{one-to-one} mapping setting, building a neural model for emulating the behavior of only \emph{one} device (i.e., a pedal or an amp) at a time
\cite{wright2021neural, wright2023adversarial}. 
Some models make further simplification and model only a certain parameter setting (a.k.a., ``snapshot'') of a device and do not take any condition signals \cite{wright2020real,martinez2020deep,comunita2023modelling}, 
while others 
incorporate device parameters as conditions for a ``full'' modeling of the device  \cite{wright2019real, steinmetz2021efficient}.

Little work, if any, has been done to tackle the more challenging task of building a single universal model that can emulate multiple devices at once.
This is harder as different devices are built from varying combinations and configurations of analog circuits, leading to distinct sonic characteristics.
We posit that a transition to such a \emph{one-to-many} 
setting is beneficial.
On one hand, it holds the key towards neural audio effect modeling with broader versatility, better reflecting the complexity seen in the real world.
On the other hand, doing so may empower the single model to learn the similarities and distinctions among different audio effects, building up a ``tone space'' that permits interpolation and extrapolation of \emph{seen} tones (i.e., tones made available to the model during training time) to approximate an \emph{unseen} tone or to create a \emph{new} tone.\footnote{We note that, the idea of using ``devices'' to define tones is less applicable in the neighboring automatic mixing tasks \cite{martinez2022automatic, koo2023music}.} 

We are in particular interested in the case of emulating unseen tones in this paper. 
Specifically, we envision a \emph{zero-shot} scenario where we have a model that can take a reference audio signal in the style of an unseen tone as the input condition, and learn to \emph{clone} the tone on-the-fly during the inference time, with no (i.e., ``zero'') model re-training. 
See Figure \ref{fig:intro} for an illustration.

In this paper, 
we set forth to tackle 
one-to-many neural modeling 
of multiple guitar amps by a single model.
Our training data contains pairs of clean (dry) signal and the corresponding wet signal rendered by an amp, out of $N=9$ possible guitar amps featuring low-gain and high-gain ones. 
Instead of building $N$ models, one for each amp, our one-to-many model can take a condition signal indicating the specific amp tone of interest and convert a given clean signal into the corresponding wet signal at inference time.
The attempt to build such an end-to-end one-to-many model represents the first contribution of this work.
 
\begin{figure}[t]
 \centerline{{
 \includegraphics[width=1.0\columnwidth]{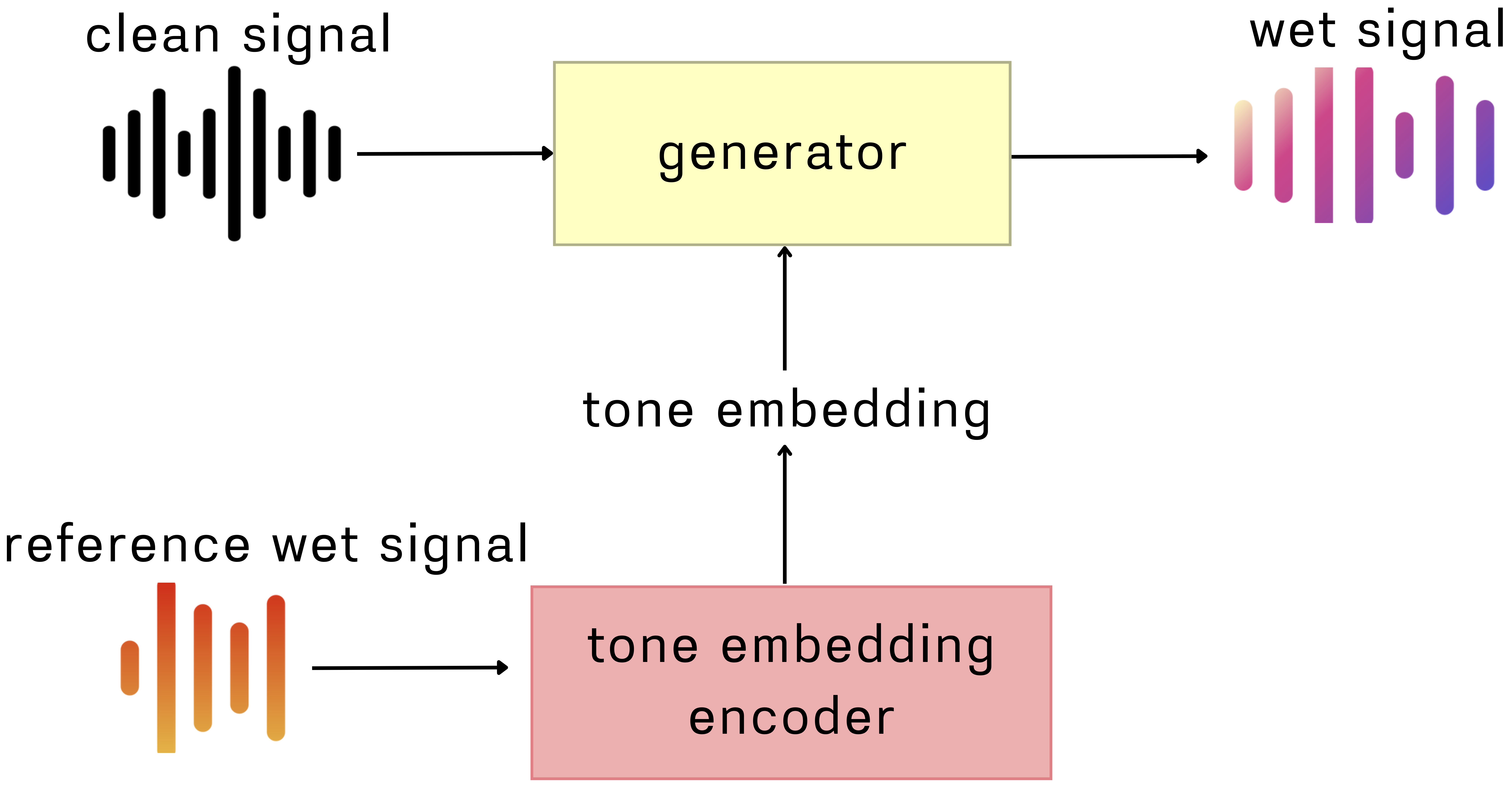}}}
 \caption{Diagram of the audio processing workflow. A clean signal $\mathbf{x}$ is input into the generator $\mathcal{G}$, which uses the tone embedding from the tone embedding encoder to produce the wet signal $\mathbf{y}$. The encoder $\mathcal{E}$ generates the tone embedding $\phi$ by analyzing a reference wet signal $\mathbf{z}$.}
 \label{fig:models_flow}
\end{figure}

While a straightforward approach to condition the generation process is via using a look-up table (LUT) to indicate the target amplifier, 
the LUT approach cannot deal with unseen amplifiers.
Targeting zero-shot applications, we use contrastive learning \cite{chen2020simple} to build a \emph{tone embedding encoder} to capture tone- (or style-) related information of a referenced audio, and then use the resulting tone embedding 
as the condition for generation,
as illustrated in Figure \ref{fig:models_flow}. 
We show via experiments that, for seen guitar amps, the tone embedding seems information richer than the LUT embedding, leading to more effective one-to-many modeling. Moreover, the tone embedding approach also works well for zero-shot learning of unseen amps, as we can turn arbitrary referenced audio of that amp into a condition to clone its style. 
In our evaluation, we use not only two additional guitar amps not seen during training time, but also self-recorded audio signals of guitar playing to validate the model's effectiveness in real-world conditions. 
The idea to realize zero-shot tone transfer 
in our one-to-many model stands as the second contribution of our work.


We invite readers to visit our demo website 
for audio samples demonstrating the result of our model.\footnote{\url{https://ss12f32v.github.io/Guitar-Zero-Shot/}}

\section{Background}


The process of applying effects to an audio signal in the real world can be described by $\mathbf{y} = f(\mathbf{x}, \phi)$, where $\mathbf{x} \in \mathbb{R}^{C \times T}$ denotes the input signal with $C$ channels and $T$ samples, $\mathbf{y} \in \mathbb{R}^{C' \times T}$ the processed output, and $\phi \in \mathbb{R}^M$ the control signal with $M$ distinct control parameters.  
The function $f$ encapsulates 
the accumulative transformation exerted by the devices involved in the effect chain connecting the input and the output.
Neural audio effect modeling aims to replicate this implicit function $f$ using either a single neural network \cite{martinez2020deep,martinez2022automatic} (i.e., the ``end-to-end'' approach) or a cascade of modularized networks~\cite{steinmetz2021automatic,steinmetz2022style}.
The latter approach is feasible only when the devices involved in the transformation is known beforehand, not applicable to the envisioned zero-shot scenario. Therefore, we focus on the the end-to-end approach in this work. 

Common \emph{backbone} models for end-to-end audio effect modeling include 
convolution neural network (CNN)-based \cite{wright2020real, steinmetz2021efficient, damskagg2019deep, comunita2023modelling}, recurrent neural network (RNN)-based \cite{wright2019real, juvela2023end}, and hybrid models \cite{martinez2022automatic, martinez2020deep}.
As mentioned in Section~\ref{sec:intro}, existing models mostly model one device at a time, sometimes even neglecting the control parameters $\phi$. For those models that consider control parameters, a ``condition representation'' to represent the control signal and a ``conditioning mechanism'' to condition the generation process by the control signal is needed.

For the \emph{condition representation}, the common approach in the field of neural audio effect modeling  is via quantizing the device parameters and then using a one-hot encoding or look-up table (LUT) to indicate the specific parameter setting of interest. 
Such \emph{ID-based} embeddings has also been used in other one-to-many audio modeling tasks, such as singing voice conversion (SVC)~\cite{deng2020pitchnet, takahashi2021hierarchical, liu2021fastsvc}.
Importantly, the LUT embeddings 
are learned at training time and then fixed at inference time, thereby failing to accommodate unseen conditions.
This is not a problem for modeling the parameters of a device, because there will not be unseen parameters for a known device.
However, it is an issue in our one-to-many setting, as we are not able to exhaustively include all the possible devices at training time.

For the \emph{conditioning mechanism}, a common approach is via ``concatenation,'' expanding $\phi$ over time to get $\phi^+ \in \mathbb{R}^{M \times T}$ and concatenating it with the input $\mathbf{x}$ channel-wisely, forming a new input $\mathbf{x}^+ \in \mathbb{R}^{(C+M) \times T}$ to the model.
Concatenation has been used by both CNN- \cite{damskagg2019deep} and RNN-based models \cite{wright2019real}.
Another common approach is via ``feature-wise linear modulation'' (FiLM) \cite{perez2018film}, which has usually been adopted by CNN-based models \cite{steinmetz2021efficient, yin2024modeling}.
FiLM injects conditions to the model by using $\phi \in \mathbb{R}^{M}$ as the input to predict different scaling $\gamma_l^c$ and shifting $\beta_l^c$ coefficients through a few linear (dense) layers, and then performing element-wise affine transformation of the intermediate feature maps of each layer of the backbone model, i.e., $\text{FiLM}(\mathbf{F}_l^c, \gamma_l^c, \beta_l^c) = \gamma_l^c \mathbf{F}_l^c + \beta_l^c$, for each layer $l$ and each $c$-th channel of the corresponding feature map.

Our work differs from the prior work  in the following three aspects.
First, we tackle multi-device modeling, conditioning our network by ``devices'' rather than ``parameters'' of a single device.
Second, we adopt the idea of \emph{content-based} embeddings 
developed in SVC \cite{zhang2020durian,guo2020phonetic,guo2022improving,li2022hierarchical,wu22ismir} to compute the condition representation instead of the ID-based embeddings, so as to deal with unseen devices. 
This is specifically done via the proposed tone embedding encoder, whose details are introduced in Section \ref{method_tone_emb}. Finally, we report experiments on zero-shot tone transfer.

\section{Multi-Tone Amplifier Modeling}


Our goal is to develop a conditional generator, denoted as $\mathcal{G}(\mathbf{x}, \phi)$, that can replicate the effect of an amplifier's audio effect chain $f$, 
guided by a tone embedding $\phi$. The objective of the generator $\mathcal{G}$ is to produce an output $\mathbf{y}$ from the input  $\mathbf{x}$, where $\mathbf{x}$ and $\mathbf{y}$ are temporally aligned and share the same musical content, but exhibit different tones (i.e., clean tone vs. a target amp tone). In our work, the tone embedding $\phi$ is a learnable representation of various tones.
Specifically, as depicted in Figure \ref{fig:models_flow}, we employ a  
tone embedding encoder, denoted as $\mathcal{E}$, to derive the tone embedding $\phi$ from a reference audio signal $\mathbf{z}$, with $\phi = \mathcal{E}(\mathbf{z})$. It is important to note that the reference signal $\mathbf{z}$ and the target $\mathbf{y}$ \emph{must} match in tone, but their musical content \emph{can} be different. More details on this in Section \ref{method_tone_emb_source}.



\subsection{Tone Embedding Encoder}
\label{method_tone_emb}


Our goal is to train an encoder $\mathcal{E}$ that can extract tone- (or style-) related features from a given wet guitar audio signal, while neglecting content-related information. Namely, this entails style/content disentanglement.
We propose to employ the self-supervised contrastive learning framework of SimCLR \cite{chen2020simple}, which was originally  for images \cite{caron2020unsupervised,assran2022masked}, to train such an audio encoder.
Specifically, our idea is to treat pairs of audio clips with \emph{different} playing contents but the \emph{same} tone as the ``positive'' pairs, and otherwise the ``negative'' pairs. Any audio clip would go through the same audio encoder to get an embedding representation of that clip, and the learning objective of SimCLR is to train the encoder such that the embeddings of clips from a positive pair are close to each other, while embeddigns of clips from a negative pair are separated apart.
By virtue of the way positive and negative data pairs is constructed, the encoder learns to project clips of the same tone to similar places in the embedding space, regardless of the underlying musical content.\footnote{The use of contrastive learning for representation learning has been widely done in the musical audio domain before, for tasks such as music classification \cite{spijkervet21ismir,gong2021ast,9763018,zhao2022s3t} and automatic mixing \cite{koo2023music}. 
What's different
here is the application of contrastive learning to the domain of guitar amp tones (which has not been attempted before, to our best knowledge), and the way we prepare positive/negative pairs for tone representation learning.}

It is easier to train the encoder if we have a collection of realistic wet guitar signals featuring different combinations of contents and tones.
Such a dataset, however, is hard to come by. 
Guitar signals collected in the wild need to be transcribed and labeled to get content- and style-related information. 
Alternatively, we may use software simulation to convert a set of clean signals into different wet signals, but the tones supported by open-source tools such as Pedalboard \cite{rice2023general,sobot_peter_2023_7817838} are limited in diversity.
Through a joint work with Positive Grid, a leading guitar amp and effect modeling company, we have the advantage of accessing high-quality and diverse data.
The training data for the encoder $\mathcal{E}$ 
is from using
a larger number of 
clean guitar signals as input to the company's commercial software to render wet signals with 
a great diversity of tones using different combinations of amplifiers and effect pedals.

\begin{figure}[t]
 \centerline{{
 \includegraphics[width=0.6\columnwidth]{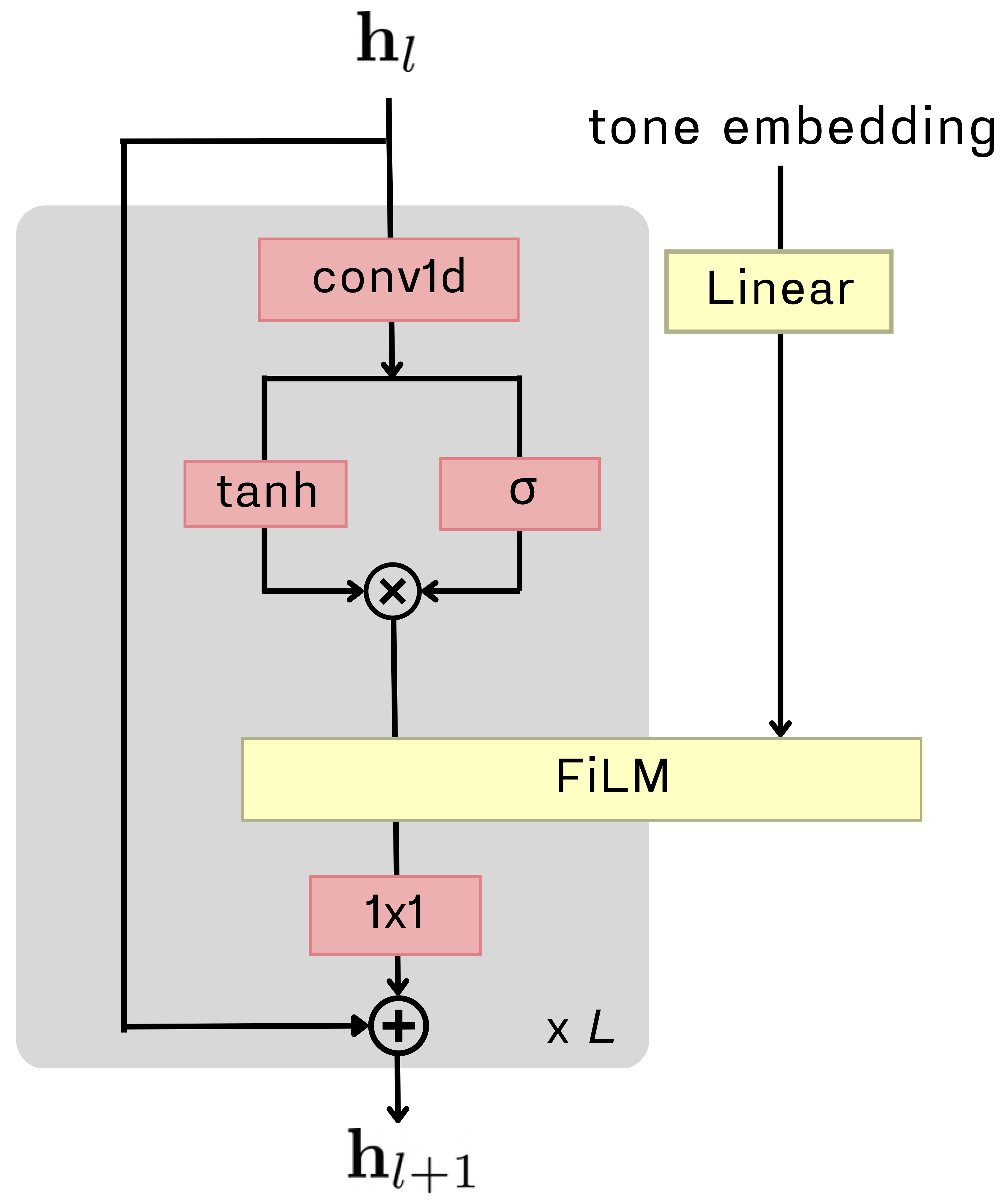}}}
 \caption{Diagram of a layer of the generator $\mathcal{G}$, which uses gated convolutional neural network (GCN)~\cite{comunita2023modelling} as the  backbone and FiLM~\cite{perez2018film} for conditioning; $\mathbf{h}_{l}$ denotes the output of the previous layer and $\mathbf{h}_{l+1}$ the current layer.}
 \label{fig:model}
\end{figure}

\subsection{Conditional Generator}

Paying more attention on the condition representation part (i.e., $\phi$ and $\mathcal{E}$), we use existing methods for the model backbone and the conditioning mechanism for our conditional generator $\mathcal{G}$. Specifically, we adapt the gated convolutional neural network (GCN)~\cite{comunita2023modelling} as our model backbone, for its demonstrated efficacy in neural audio effect modeling.
For the conditioning mechanism, we use FiLM~\cite{perez2018film}. 
As shown in Figure \ref{fig:model}, each layer of GCN passes the output of the previous layer $\mathbf{h}_l$ through a 1D convolution with a progressively increasing dilation factor, followed by sigmoid and tanh for gated activation. 
The resulting feature map is then modulated by the FiLM module, before 
being processed by a 1x1 conv1d and then added with the output of the previous layer with a residual link. The same tone embedding $\phi$ is used in all the $L$ layers of GCN, but each layer $l$ has its own learnable parameters. 
The outputs from all layers are finally concatenated and processed through a final 1x1 conv1d mixing layer to generate the  output signal $\mathbf{y}$.\footnote{Unlike \cite{comunita2023modelling}, we pad zeros at the start of the signal but not in the intermediate feature maps, for otherwise there would be unwanted impulse-like sounds. This means each layer ends up being shorter in our case. To keep the output size consistent across layers, we crop the residual part of each layer following the causal principle.}


The tone embedding encoder $\mathcal{E}$ is trained with a large set of wet signals rendered with different combinations of amps and effect pedals,  while the generator $\mathcal{G}$ is trained separately with a smaller set of paired data of clean and wet signals rendered with different amps (not using effect pedals). The encoder $\mathcal{E}$ is trained beforehand and then fixed (i.e., parameters frozen) while training the generator $\mathcal{G}$.

The training data of $\mathcal{G}$ is also from Positive Grid, containing 30 minutes of clean input data (for $\mathbf{x}$) rendered with 9 different guitar amplifiers (for $\mathbf{y}$). 
For performance evaluation, we divide the clean signals into training, validation, and test sets with an 80/10/10 ratio.
According to the company's taxonomy, 
there are 3 types of amps:
\begin{itemize}[leftmargin=*,itemsep=0pt,topsep=2pt]
\item \emph{High-gain} tones are perceived as highly distorted. We have Mesa Boogie Mark IV (amp1), PRS Archon 100 (amp2), and Soldano SLO-100 (amp3). 
\item \emph{Low-gain} tones are often recognized as an overdrive sound. Our data contains Fender Tweed Deluxe (amp4), Vox AC30 (amp5), and Matchless DC30 (amp6). 
\item \emph{Crunch} tones exhibit a mid-range gain level, in between low- and high-gain.  Our data contains Vox AC30 Handwired Overdriven (amp7), Friedman BE100 (amp8), and Overdriven Marshall JTM45 (amp9).
\end{itemize}

\subsection{Source of the Reference Audio Signal}
\label{method_tone_emb_source}


While training $\mathcal{G}$, for each data pair $\{\mathbf{x},\mathbf{y}\}$, the encoder $\mathcal{E}$ takes a reference signal $\mathbf{z}$ providing information of the tone of $\mathbf{y}$. Therefore, $\mathbf{y}$ and $\mathbf{z}$ must match in tone. However, interestingly, $\mathbf{y}$ and $\mathbf{z}$ can be different in content. 

The na\"ive \textbf{paired reference} method of simply setting $\mathbf{z}=\mathbf{y}$ demands the reference signal and target output, and accordingly the input $\mathbf{x}$, to play the same  content. This is fine at training time, but is too restrictive and not practical at inference time, especially for zero-shot scenarios. 
Following the idea of using ``target-unaligned audio'' of a prior work \cite{chen24dafx}, we instead employ an \textbf{unpaired reference} method, selecting a signal $\mathbf{z}$ \emph{at random} from the training set as long as it is rendered with the same amp tone as $\mathbf{y}$.

Besides zero-shot capability, the proposed unpaired reference method may further encourage style/content disentanglement, because here $\phi$ only provides information concerning the tone of the target, not its content.

\subsection{Zero-shot Tone Transfer}
\label{method_zero_shot}

At inference time, the proposed model $\mathcal{G}$ holds the potential to clone the tone of a reference signal $\mathbf{z}^*$ for even unseen tones and unseen reference signals, via using $\mathcal{E}(\mathbf{z}^*)$ as the condition $\phi^*$.
This might be feasible as the encoder $\mathcal{E}$ has actually been trained on a great diversity of tones besides the limited number of $N$ tones used to train $\mathcal{G}$.

Besides using $\phi^*=\mathcal{E}(\mathbf{z}^*)$, we consider the following two \emph{retrieval-based} alternatives to get $\phi^*$, 
referred to as ``nearest-embedding'' and ``mean-embedding'' respectively.
\begin{itemize}[leftmargin=*,itemsep=0pt,topsep=2pt]

    \item $\phi^*= \underset{\phi \in \Phi}{\arg\max} ~\text{sim}(\mathcal{E}(\mathbf{z}^*),\phi$), where $\phi=\mathcal{E}(\mathbf{z})$ denotes the embedding for a reference signal \emph{seen} and sampled from the training set, $\text{sim}(\cdot,\cdot)$ the cosine similarity between two vectors, and $\Phi$ the collection of embeddings for such seen reference signals from the training set to be compared against the query $\mathcal{E}(\mathbf{z}^*)$. Namely, this method picks the known reference signal $\mathbf{z}$ whose embedding is closest to that of the unseen reference $\mathbf{z}^*$ as the surrogate to condition the generation process. We set the size of the candidate set $|\Phi|=3,600$ in our implementation.
    
    \item $\phi^*= \underset{\phi \in \{\psi_1,\psi_2\,\dots,\psi_N\} }{\arg\max} ~\text{sim}(\mathcal{E}(\mathbf{z}^*),\phi$), where $\psi_n$ stands for the average of the embeddings associated with the $n$-th amp tone (out of the $N$ seen amp tones) from the aforementioned candidate set $\Phi$. 
    This can be viewed as an LUT-like approach because we use one mean embedding to represent each amp tone for retrieval.

    
 
\end{itemize}


\subsection{Implementation Details}

The encoder $\mathcal{E}$ is developed by Positive Grid using in-house data and implementation. 
It is an audio encoder 
that processes mel-spectrograms of guitar signals, 
trained with a batch size of 200 
short audio clips sampled at 16kHz, along with data augmentation techniques such as adding noise and random cropping.

For the generator $\mathcal{G}$, we followed \cite{steinmetz2022style} and applied --12 dBFS peak normalization to the training data to balance the sound levels of different amplifiers and ensure headroom for distortion. 
We randomly paired a 3.5-second monaural clean input sampled at 44.1kHz with an amp output from the 9 amplifiers as the training examples. 
We trained the generator $\mathcal{G}$ 
on an NVIDIA RTX 3090 GPU (with 24 GB memory), using the Adam optimizer \cite{KingBa15} with a learning rate of 1e--3 and a batch size of 12. 
While both the input and output of $\mathcal{G}$ are time-domain waveforms, we used complex-valued spectral loss as the training objective, with an STFT window length of 2,048 and a hop length of 512, for this loss function led to 
better result in our pilot study.


For the architecture of $\mathcal{G}$, we configured the GCN with $L=12$ conv1D layers, each with 16 channels, followed by a final 1x1 conv1d layer combining the outputs of all the preceding layers to a monaural waveform. For conditioning, we firstly projected each 512-dim tone embedding produced by $\mathcal{E}$ to a 128-dim vector, then applied a series of 10 linear layers at each GCN layer to predict the scaling and shifting coefficients $\gamma_l^c$ and  $\beta_l^c$  for FiLM.
The total number of trainable parameters of $\mathcal{G}$ is around 120k, and the model training of $\mathcal{G}$ converged in 1/2 days.

\newcommand{\testmode}[2] {%
    \begin{tabularx}{@{}c@{}}
        \textbf{#1} \\ \textbf{d=#2}
    \end{tabularx}%
}

\begin{table*}[t]
\centering
\begin{tabular}{clc|ccc|cc}
\toprule
&& GCN & \multicolumn{3}{c|}{FiLM-GCN} & \multicolumn{2}{c}{Concat-GCN}  \\
\cmidrule(lr){3-3} \cmidrule(lr){4-6} \cmidrule(lr){7-8}
&& one-to-one & LUT & ToneEmb\,(paired)  & ToneEmb\,(unpaired)   & LUT  & ToneEmb\,(paired)  \\
\midrule
&{Amp1} & 0.0420& 0.1441 & 0.1189& \textbf{0.0777} & 0.1593  & 0.1523  \\
high-gain &{Amp2} & 0.0268 & 0.1951 & \textbf{0.0670} & 0.1189 & 0.1741  & 0.1208  \\
&{Amp3} & 0.0527  & 0.1659 & 0.1254&\textbf{0.1143} & 0.1777  & 0.1304  \\
\midrule
&{Amp4} & 0.0087& 0.0698 & \textbf{0.0230}& 0.0275  & 0.0618  & 0.0775  \\
low-gain &{Amp5} & 0.0004 & 0.0813 & 0.0167&\textbf{0.0138}  & 0.0334  & 0.0166  \\
&{Amp6} & 0.0014   &  0.0947 & 0.0169&\textbf{0.0121} &  0.0779  &0.0275 \\
\midrule
&{Amp7} & 0.0393  & 0.1022 & 0.0860&0.0885 & \textbf{0.0733}  & 0.0988 \\
crunch &{Amp8}  & 0.0124 & 0.1583 & 0.0760&\textbf{0.0604} & 0.1562  & 0.0775 \\
&{Amp9}  & 0.0035  & 0.1593 & 0.0375&\textbf{0.0290}& 0.1211  & 0.0407 \\

\bottomrule
\end{tabular}
\caption{Efficacy of various models in modeling 9 guitar amplifiers, measured in complex STFT loss on the test data. The leftmost column shows the result of the baseline \emph{one-to-one} GCN approach, building one model per amp. The others are \emph{one-to-many}, building a single model for all the 9 amps, using  FiLM (middle three) or concatenation (last two) as the conditioning mechanism.
For conditioning representation, we evaluate LUT and the proposed tone embedding (`ToneEmb') with either paired  or unpaired reference (cf. Section \ref{method_tone_emb_source}). Best results of one-to-many models are highlighted.}
\label{tab:model_results}
\end{table*}

\begin{figure}[h]
 \centerline{{
 \includegraphics[width=0.8\columnwidth]{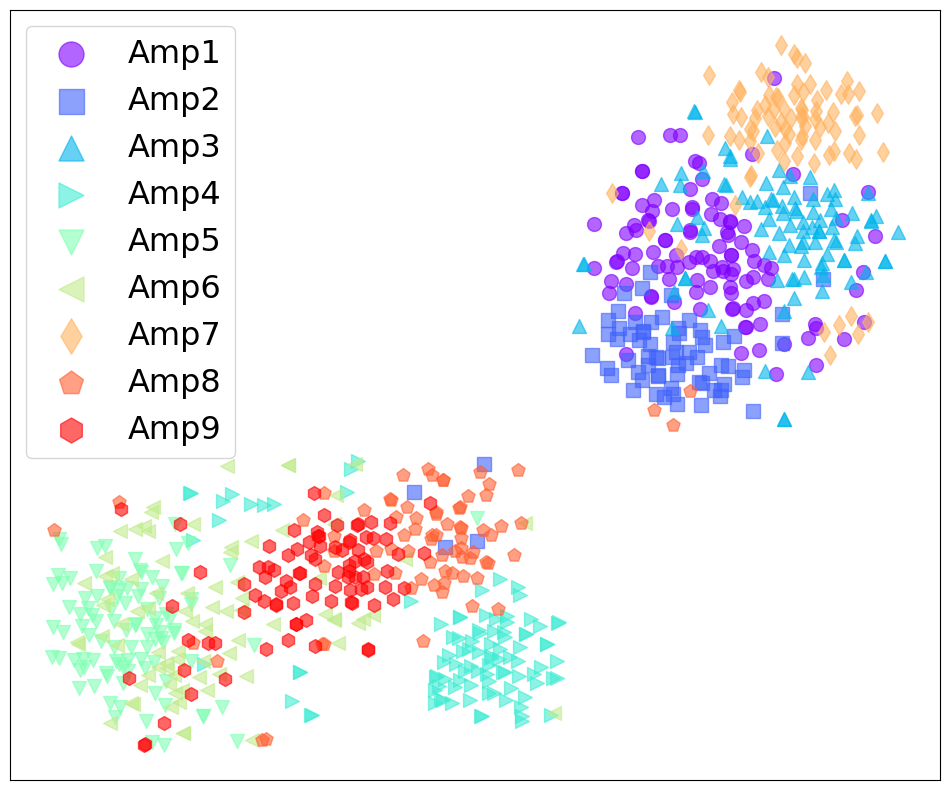}}}
 \caption{A t-SNE visualization of the tone embeddings from the wet signals of the $N=9$ amps.
 Each point represents a tone embedding extracted from a wet signal, with color and shape indicating the category of the amp tone. 
 We see 2 big cross-amp clusters and 9 small clusters for each amp,  suggesting the ability of the encoder $\mathcal{E}$ to distinguish between different tones based on their embeddings.}
 \label{fig:tsne}
\end{figure}

\section{Experimental Results}

\subsection{Tone Embedding Visualization}

We examine the tone embedding space of the 9 target amps first, using t-SNE \cite{van2008visualizing} for dimension reduction and visualization. 
Figure \ref{fig:tsne} shows that the  embeddings of the wet signals of the same amp, while differing in musical content, indeed cluster together in the projected 2-D space, demonstrating the efficacy of our encoder $\mathcal{E}$ in capturing tone-related information.
Moreover, we see two big clusters, which somehow separate high-gain amps (amps 1--3) from low-gain amps (amps 4--6).
There are also overlaps, suggesting some similarity among the amps.
Amp 7 is closer to high-gain, while amps 8 and 9 closer to low-gain.

\subsection{Efficacy of One-to-many Neural Amp Modeling}
\label{comparison_emebdding_sources}
Next, we evaluate the efficacy of modeling the 9 target amps, comparing different approaches (one-to-one versus one-to-many), conditioning mechanisms (FiLM or the simpler concatenation), and condition representation (the proposed tone embedding, `ToneEmb' for short, or LUT). 
We implement all models using GCN as the backbone.

Result shown in Table \ref{tab:model_results} leads to several observations. 
From the leftmost column, we see low-gain amps seem easier to model than crunch amps, while high-gain amps are the most challenging (i.e., with higher testing losses). 
Signals from a high-gain amp are highly-distorted, with more high-frequency components that may be harder to be modeled.
Similar trends can be seen from other columns.

The middle columns show the result of the one-to-many GCN with FiLM conditioning (`FiLM-GCN'). 
Firstly, we see the losses are in general higher than those of the one-to-one non-conditional baseline. 
This is expected, as we treat the result of the one-to-one model as a performance upperbound, for one-to-one modeling is inherently easier.
Among the three variants of FiLM-GCN, ToneEmb with unpaired reference leads to the best result for most amps, reducing greatly the performance gap between the one-to-one approach and the variant with the straightforward LUT-based conditioning representation.
While LUT only learns $N$ unique embeddings $\phi$, one for each amp, ToneEmb is much more versatile as it computes a unique embedding for each reference wet signal.  It seems that the ToneEmb embeddings are thereby information richer, benefiting multi-amp modeling.

For ToneEmd, we initially expect that the advantage of `unpaired reference' over `paired reference' is on zero-shot learning of unseen amps. For seen amps,  paired reference simply uses the target wet signal $\mathbf{y}$ as the reference signal $\mathbf{z}$, providing direct and potentially stronger condition signals. To our surprise, while both `ToneEmd (paired)' and `ToneEmd (unpaired)' greatly outperform the LUT approach, 
the unpaired approach slightly outperforms the paired approach for most amps. This may be due to the stronger incentive of style/content disentanglement induced by unpaired referencing, as discussed in Section \ref{method_tone_emb_source}, but more empirical studies (which we leave as future work) are needed to confirm this.

Finally, Table \ref{tab:model_results}  shows that the ablated version of using concatenation as the conditioning mechanism  leads to worse results most of the time. We hence use FiLM-GCN trained with unpaired referencing in experiments below.

\subsection{Zero-shot Learning on Unseen Amplifiers}
\label{exp_sec:zero_shot}

To investigate the potential of the proposed methodology on zero-shot learning, we create wet signals using two ``unseen'' amplifiers---\textit{High Gain EL34 V2} (high gain) and \textit{Dumble ODS 50} (low gain)---using the clean signals from the test set, and use them as the reference signals $\mathbf{z}^*$ for FiLM-GCN, to see whether it can learn the tones zero-shot.
Namely, both the content and style are unseen at training time.
Here, we evaluate the non-retrieval-based method of using $\phi^*=\mathcal{E}(\mathbf{z}^*)$ and the two retrieval-based methods introduced in Section \ref{method_zero_shot}.

Table \ref{tab:r_vs_direct} shows that the non-retrieval-based method slightly outperforms the other two for the unseen high-gain amp, while nearest-embedding performs the best for the unseen low-gain amp.
More importantly, comparing the losses tabulated in Tables \ref{tab:model_results}  and \ref{tab:r_vs_direct}, we see that the loss for the  unseen low-gain amp is not greatly larger than the loss for the seen low-gain amps, only 1--2 times larger. The loss of the unseen high-gain amp is a bit high, but is  only about 2 times larger than those of the seen high-gain amps.
We take this as a positive indication of the efficacy of the proposed model in dealing with unseen tones.




Table \ref{tab:r_vs_direct} also shows that mean-embedding  performs the worst, adding support of using more versatile embeddings for conditioning.
In this regard, the non-retrieval-based method is actually more flexible than nearest-retrieval, as it can compute the reference embedding $\phi^*$ on-the-fly without referring to a pre-computed presumably large collection of embeddings.
Future work can be done to study its effectiveness with more amps (i.e., larger $N$).

\begin{table}[t]
\centering
\begin{tabular}{lccc}
\toprule
& non-retrieval & \multicolumn{2}{c}{retrieval-based}  \\
\cmidrule(lr){3-4} 
& ($\phi^*=\mathcal{E}(\mathbf{z}^*)$) & nearest & mean  \\
\midrule
unseen high gain & \textbf{0.2511} & 0.2560 & 0.2593  \\
unseen low gain  & 0.0338 & \textbf{0.0274} & 0.0404 \\
\bottomrule
\end{tabular}
\caption{Efficacy of using different methods for FiLM-GCN (cf. Section \ref{method_zero_shot}) for zero-shot modeling of two unseen amps, measured again in complex STFT loss.} 
\label{tab:r_vs_direct}
\end{table}

\subsection{Case Study on Zero-shot Amp Tone Transfer}
\label{exp_sec:case_study}

To further study the zero-shot scenario, we present finally a case study employing 
a self-recorded (by one of the authors) guitar solo audio signal with content and tone both unseen at the training time. 
This recording was captured using a Boss GT-1000 effects processor with a default factory preset based on a high-gain Marshall amplifier setting. The effect chain included not only this unseen amp but also an equalizer (EQ) with a high-cut filter at 10kHz.

\begin{figure}[t]
 \centerline{{
 \includegraphics[width=0.95\columnwidth]{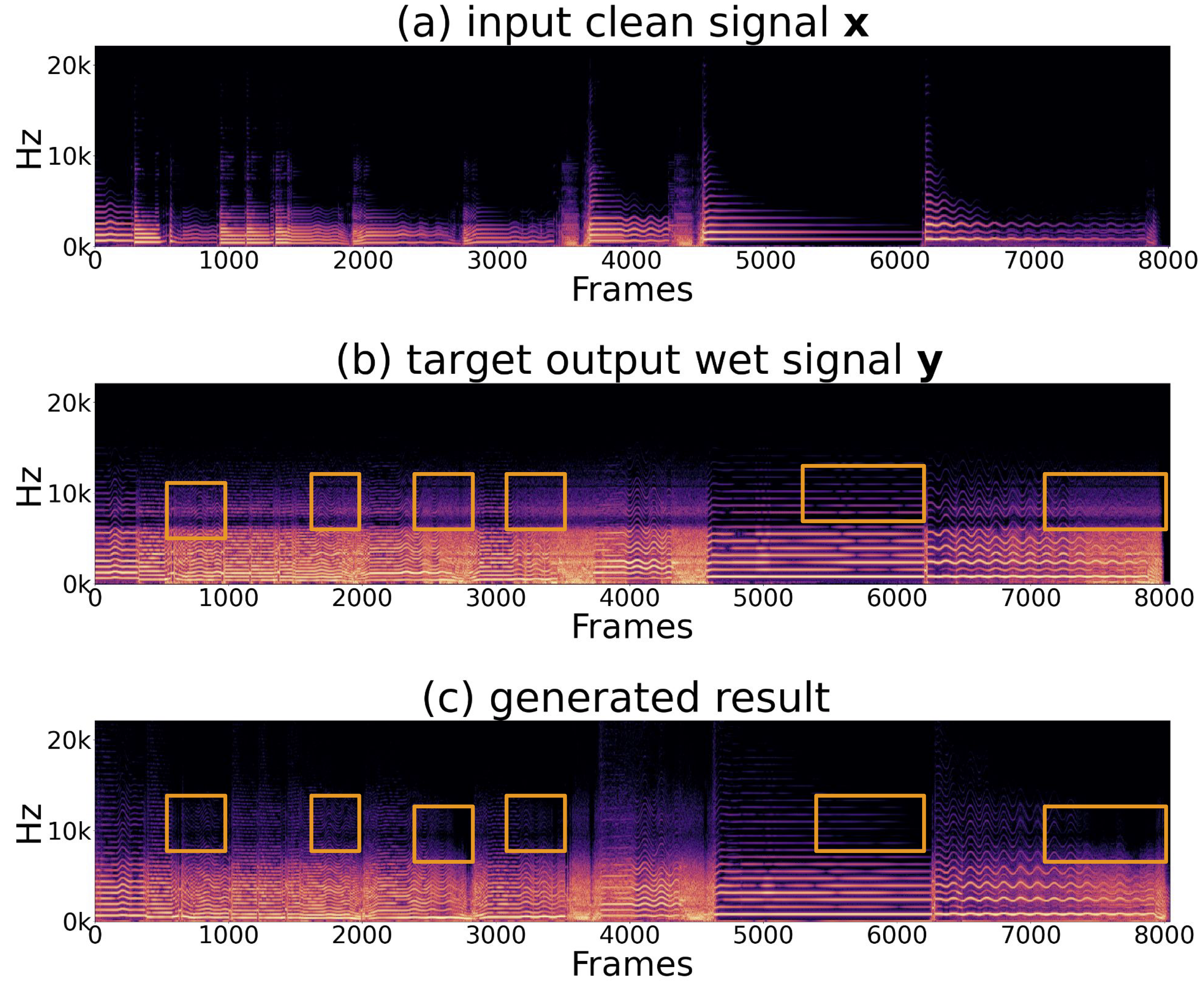}}}
 \caption{Spectrograms of the input clean signal,  target wet signal, and the generated result of the proposed one-to-many FiLM-GCN model, in the zero-shot case study reported in Section \ref{exp_sec:case_study}. The orange squares show that our  model still struggles to model high-frequency components.}
 \vspace{-2mm}
 \label{fig:zeroshot}
\end{figure}

The visualization of the spectrograms shown in Figure \ref{fig:zeroshot} suggests that the generated result possesses characteristics similar to those of the target wet signal, but there are notable difference in the high-frequency area. Since we do not consider EQ as a modeling target, our model cannot produce filter-based effects on the signal, resulting in additional harmonics in the spectrum. Furthermore, the sustain of each note is not perfectly reproduced, as the harmonics in the highlighted orange squares are not sequentially connected compared to the case of the target, indicating a struggle to model the high-frequency content accurately.

Despite these limitations, our model can still generate reasonable harmonics according to the input. For the quick string-bending content around frames 6,000 to 7,000, the generated harmonics are correctly damped. 
Our tone embedding encoder recognizes that the tone of the reference signal is closer to high-gain, empowering the generator to process the input accordingly.
We provide audio samples in the supplementary material, including a multi-track recording and a remixed audio created by rendering multiple track separately using our model.

\section{Conclusion}
In this paper, we have presented an end-to-end one-to-many methodology that uses conditions from a tone embedding encoder to emulate multiple guitar amps through a single model, providing empirical evidences of its potential for zero-shot amp modeling. Moving forward, several avenues for future work emerge. First, for more comprehensive audio effect modeling, we can apply our methodology with various configurations of audio effect chains. 
Second, we can further improve the model's effectiveness on one-to-many modeling by incorporating more advanced architectures and conditioning mechanisms, such as hyper-network-based conditioning \cite{2021hyeprconv,yeh24dafx}.
Finally, for more effective zero-shot tone transfer,
we can train the model on a wider range of amplifier types, 
which might also
pave the way for universal amplifier modeling. 

\section{Acknowledgements}
We thank Positive Grid for assistance with the datasets, guitar effect plugins, and for sharing the tone embedding model mentioned in Section \ref{method_tone_emb}. This work is also partially supported by a grant from the National Science and Technology Council of Taiwan (NSTC\,112-2222-E-002-005-MY2).

\bibliography{ISMIR}

\begin{thebibliography}{10}
\providecommand{\url}[1]{#1}
\csname url@samestyle\endcsname
\providecommand{\newblock}{\relax}
\providecommand{\bibinfo}[2]{#2}
\providecommand{\BIBentrySTDinterwordspacing}{\spaceskip=0pt\relax}
\providecommand{\BIBentryALTinterwordstretchfactor}{4}
\providecommand{\BIBentryALTinterwordspacing}{\spaceskip=\fontdimen2\font plus
\BIBentryALTinterwordstretchfactor\fontdimen3\font minus \fontdimen4\font\relax}
\providecommand{\BIBforeignlanguage}[2]{{%
\expandafter\ifx\csname l@#1\endcsname\relax
\typeout{** WARNING: IEEEtran.bst: No hyphenation pattern has been}%
\typeout{** loaded for the language `#1'. Using the pattern for}%
\typeout{** the default language instead.}%
\else
\language=\csname l@#1\endcsname
\fi
#2}}
\providecommand{\BIBdecl}{\relax}
\BIBdecl

\bibitem{damskagg2019deep}
E.-P. Damsk{\"a}gg, L.~Juvela, E.~Thuillier, and V.~V{\"a}lim{\"a}ki, ``Deep learning for tube amplifier emulation,'' in \emph{IEEE International Conference on Acoustics, Speech and Signal Processing (ICASSP)}, 2019.

\bibitem{wright2019real}
A.~Wright, E.-P. Damsk{\"a}gg, and V.~V{\"a}lim{\"a}ki, ``Real-time black-box modelling with recurrent neural networks,'' in \emph{International Conference on Digital Audio Effects}, 2019.

\bibitem{wright2020real}
A.~Wright, E.-P. Damsk{\"a}gg, L.~Juvela, and V.~V{\"a}lim{\"a}ki, ``Real-time guitar amplifier emulation with deep learning,'' \emph{Applied Sciences}, vol.~10, no.~3, p. 766, 2020.

\bibitem{martinez2020deep}
M.~A. Mart{\'\i}nez~Ram{\'\i}rez, E.~Benetos, and J.~D. Reiss, ``Deep learning for black-box modeling of audio effects,'' \emph{Applied Sciences}, vol.~10, no.~2, p. 638, 2020.

\bibitem{juvela2023end}
L.~Juvela, E.-P. Damsk{\"a}gg, A.~Peussa, J.~M{\"a}kinen, T.~Sherson, S.~I. Mimilakis, K.~Rauhanen, and A.~Gotsopoulos, ``End-to-end amp modeling: from data to controllable guitar amplifier models,'' in \emph{IEEE International Conference on Acoustics, Speech and Signal Processing (ICASSP)}, 2023.

\bibitem{comunita2023modelling}
M.~Comunit{\`a}, C.~J. Steinmetz, H.~Phan, and J.~D. Reiss, ``Modelling black-box audio effects with time-varying feature modulation,'' in \emph{IEEE International Conference on Acoustics, Speech and Signal Processing (ICASSP)}, 2023.

\bibitem{yin2024modeling}
H.~Yin, G.~Cheng, C.~J. Steinmetz, R.~Yuan, R.~M. Stern, and R.~B. Dannenberg, ``Modeling analog dynamic range compressors using deep learning and state-space models,'' \emph{arXiv preprint arXiv:2403.16331}, 2024.

\bibitem{chen24dafx}
Y.-H. Chen, W.~Choi, W.-H. Liao, M.~A. Mart{\'\i}nez~Ram{\'\i}rez, K.~W. Cheuk, Y.~Mitsufuji, J.-S.~R. Jang, and Y.-H. Yang, ``Improving unsupervised clean-to-rendered guitar tone transformation using {GANs} and integrated unaligned clean data,'' in \emph{International Conference on Digital Audio Effects (DAFx)}, 2024.

\bibitem{yeh24dafx}
Y.-T. Yeh, W.-Y. Hsiao, and Y.-H. Yang, ``Hyper recurrent neural network: Condition mechanisms for black-box audio effect modeling,'' in \emph{International Conference on Digital Audio Effects (DAFx)}, 2024.

\bibitem{steinmetz2021automatic}
C.~J. Steinmetz, J.~Pons, S.~Pascual, and J.~Serr{\`a}, ``Automatic multitrack mixing with a differentiable mixing console of neural audio effects,'' in \emph{IEEE International Conference on Acoustics, Speech and Signal Processing (ICASSP)}, 2021.

\bibitem{martinez2022automatic}
M.~A. Mart{\'\i}nez~Ram{\'\i}rez, W.~Liao, C.~Nagashima, G.~Fabbro, S.~Uhlich, and Y.~Mitsufuji, ``Automatic music mixing with deep learning and out-of-domain data,'' in \emph{International Society for Music Information Retrieval (ISMIR)}, 2022.

\bibitem{koo2023music}
J.~Koo, M.~A. Martínez-Ramírez, W.-H. Liao, S.~Uhlich, K.~Lee, and Y.~Mitsufuji, ``Music mixing style transfer: A contrastive learning approach to disentangle audio effects,'' \emph{arXiv preprint arXiv:2211.02247}, 2023.

\bibitem{mimilakis2020one}
S.~I. Mimilakis, N.~J. Bryan, and P.~Smaragdis, ``One-shot parametric audio production style transfer with application to frequency equalization,'' in \emph{IEEE International Conference on Acoustics, Speech and Signal Processing (ICASSP)}, 2020, pp. 256--260.

\bibitem{steinmetz2022style}
C.~J. Steinmetz, N.~J. Bryan, and J.~D. Reiss, ``Style transfer of audio effects with differentiable signal processing,'' \emph{J. Audio Eng. Soc}, vol.~70, no.~9, pp. 708--721, 2022.

\bibitem{wright2021neural}
A.~Wright and V.~Valimaki, ``Neural modeling of phaser and flanging effects,'' \emph{Journal of the Audio Engineering Society}, vol.~69, no. 7/8, pp. 517--529, 2021.

\bibitem{wright2023adversarial}
A.~Wright, V.~V{\"a}lim{\"a}ki, and L.~Juvela, ``Adversarial guitar amplifier modelling with unpaired data,'' in \emph{IEEE International Conference on Acoustics, Speech and Signal Processing (ICASSP)}, 2023.

\bibitem{steinmetz2021efficient}
C.~J. Steinmetz and J.~D. Reiss, ``Efficient neural networks for real-time analog audio effect modeling,'' in \emph{152nd Audio Engineering Society Convention}, 2022.

\bibitem{chen2020simple}
T.~Chen, S.~Kornblith, M.~Norouzi, and G.~Hinton, ``A simple framework for contrastive learning of visual representations,'' in \emph{International Conference on Machine Learning}, 2020, pp. 1597--1607.

\bibitem{deng2020pitchnet}
C.~Deng, C.~Yu, H.~Lu, C.~Weng, and D.~Yu, ``{PitchNet}: Unsupervised singing voice conversion with pitch adversarial network,'' in \emph{IEEE International Conference on Acoustics, Speech and Signal Processing (ICASSP)}, 2020, pp. 7749--7753.

\bibitem{takahashi2021hierarchical}
N.~Takahashi, M.~K. Singh, and Y.~Mitsufuji, ``Hierarchical disentangled representation learning for singing voice conversion,'' in \emph{International Joint Conference on Neural Networks (IJCNN)}, 2021, pp. 1--7.

\bibitem{liu2021fastsvc}
S.~Liu, Y.~Cao, N.~Hu, D.~Su, and H.~Meng, ``{FastSVC}: Fast cross-domain singing voice conversion with feature-wise linear modulation,'' in \emph{IEEE International Conference on Multimedia and Expo (ICME)}, 2021, pp. 1--6.

\bibitem{perez2018film}
E.~Perez, F.~Strub, H.~De~Vries, V.~Dumoulin, and A.~Courville, ``{FiLM}: Visual reasoning with a general conditioning layer,'' in \emph{AAAI Conference on Artificial Intelligence}, 2018.

\bibitem{zhang2020durian}
L.~Zhang, C.~Yu, H.~Lu, C.~Weng, C.~Zhang, Y.~Wu, X.~Xie, Z.~Li, and D.~Yu, ``{DurIAN-SC}: Duration informed attention network based singing voice conversion system,'' in \emph{International Speech Communication Association (INTERSPEECH)}, 2020, pp. 1231--1235.

\bibitem{guo2020phonetic}
H.~Guo, H.~Lu, N.~Hu, C.~Zhang, S.~Yang, L.~Xie, D.~Su, and D.~Yu, ``Phonetic posteriorgrams based many-to-many singing voice conversion via adversarial training,'' \emph{arXiv preprint arXiv:2012.01837}, 2020.

\bibitem{guo2022improving}
H.~Guo, Z.~Zhou, F.~Meng, and K.~Liu, ``Improving adversarial waveform generation based singing voice conversion with harmonic signals,'' in \emph{IEEE International Conference on Acoustics, Speech and Signal Processing (ICASSP)}, 2022, pp. 6657--6661.

\bibitem{li2022hierarchical}
X.~Li, S.~Liu, and Y.~Shan, ``A hierarchical speaker representation framework for one-shot singing voice conversion,'' \emph{International Speech Communication Association (INTERSPEECH)}, pp. 4307--4311, 2022.

\bibitem{wu22ismir}
J.-T. Wu, J.-Y. Wang, J.-S.~R. Jang, and L.~Su, ``A unified model for zero-shot singing voice conversion and synthesis,'' in \emph{International Society for Music Information Retrieval Conference (ISMIR)}, 2022.

\bibitem{caron2020unsupervised}
M.~Caron, I.~Misra, J.~Mairal, P.~Goyal, P.~Bojanowski, and A.~Joulin, ``Unsupervised learning of visual features by contrasting cluster assignments,'' \emph{Advances in Neural Information Processing Systems}, pp. 9912--9924, 2020.

\bibitem{assran2022masked}
M.~Assran, M.~Caron, I.~Misra, P.~Bojanowski, F.~Bordes, P.~Vincent, A.~Joulin, M.~Rabbat, and N.~Ballas, ``Masked siamese networks for label-efficient learning,'' in \emph{European Conference on Computer Vision}.\hskip 1em plus 0.5em minus 0.4em\relax Springer, 2022, pp. 456--473.

\bibitem{spijkervet21ismir}
J.~Spijkervet and J.~A. Burgoyne, ``Contrastive learning of musical representations,'' in \emph{International Society for Music Information Retrieval (ISMIR)}, 2021.

\bibitem{gong2021ast}
Y.~Gong, Y.-A. Chung, and J.~Glass, ``{AST: Audio Spectrogram Transformer},'' in \emph{International Speech Communication Association (INTERSPEECH)}, 2021, pp. 571--575.

\bibitem{9763018}
H.~Yakura, K.~Watanabe, and M.~Goto, ``Self-supervised contrastive learning for singing voices,'' \emph{IEEE/ACM Transactions on Audio, Speech, and Language Processing}, vol.~30, pp. 1614--1623, 2022.

\bibitem{zhao2022s3t}
H.~Zhao, C.~Zhang, B.~Zhu, Z.~Ma, and K.~Zhang, ``{S3T}: Self-supervised pre-training with {Swin Transformer} for music classification,'' in \emph{IEEE International Conference on Acoustics, Speech and Signal Processing (ICASSP)}, 2022, pp. 606--610.

\bibitem{rice2023general}
M.~Rice, C.~J. Steinmetz, G.~Fazekas, and J.~D. Reiss, ``General purpose audio effect removal,'' in \emph{IEEE Workshop on Applications of Signal Processing to Audio and Acoustics (WASPAA)}, 2023.

\bibitem{sobot_peter_2023_7817838}
P.~Sobot, ``Pedalboard,'' 2021, [Online] \url{https://github.com/spotify/pedalboard}.

\bibitem{KingBa15}
D.~Kingma and J.~Ba, ``Adam: A method for stochastic optimization,'' in \emph{International Conference on Learning Representations (ICLR)}, 2015.

\bibitem{van2008visualizing}
L.~Van~der Maaten and G.~Hinton, ``Visualizing data using {t-SNE},'' \emph{Journal of Machine Learning Research}, vol.~9, no.~11, 2008.

\bibitem{2021hyeprconv}
A.~Richard, D.~Markovic, I.~D. Gebru, S.~Krenn, G.~A. Butler, F.~Torre, and Y.~Sheikh, ``Neural synthesis of binaural speech from mono audio,'' in \emph{Proc. Int. Conf. Learning Representations}, 2021.

\end{thebibliography}

\end{document}